\documentclass[runningheads,a4paper]{llncs}

\usepackage{amssymb}
\usepackage{url}

\urldef{\mailsa}\path|brassard@iro.umontreal.ca|
\newcommand{\keywords}[1]{\par\addvspace\baselineskip
\noindent\keywordname\enspace\ignorespaces#1}

\begin{document}

\mainmatter

\title{Cryptography in a Quantum World\,\thanks{Invited talk at SOFSEM~2016.
Final publication available at \texttt{link.springer.com}.}}

\author{Gilles Brassard\,\inst{1}$\mbox{}^{,}$\inst{2}}
\authorrunning{Gilles Brassard}

\institute{D\'epartement d'informatique et de recherche op\'erationnelle\\
Universit\'e de Montr\'eal, C.P.~6128, Succursale Centre-ville\\
Montr\'eal (QC), H3C 3J7~Canada\\ \and
Canadian Institute for Advanced Research\\
\mailsa\\
\url{http://www.iro.umontreal.ca/~brassard/en/}}

\maketitle

\begin{abstract}
Although practised as an art and science for ages, cryptog\-raphy had to wait until the mid-twentieth century before Claude Shannon gave it a strong mathematical foundation. However, Shannon's approach was rooted is his own information theory, itself inspired by the classical physics of Newton and Einstein. But our world is ruled by the laws of quantum mechanics. When quantum-mechanical phenomena are taken into account, new vistas open up both for code\-makers and code\-breakers.
Is~quantum mechanics a blessing or a curse for the protection of privacy? As~we shall see, the jury is still~out!
\end{abstract}

\section{Introduction}

For thousands of years, cryptography has been an ongoing battle between code\-makers and code\-breakers~\cite{Kahn,Singh}, who are more formally called cryptographers and cryptanalysts.
Naturally, \emph{good} and \emph{evil} are subjective terms to designate code\-makers and code\-breakers.
As~a passionate advocate for the right to privacy, my allegiance is clearly on the side
of code\-makers. I~admit that I~laughed hysterically when I~saw the \emph{Zona Vigilada}
warning that awaits visitors of the \emph{Pla\c{c}a de George Orwell} near City Hall
in Barcelona~\cite{Orwell}.
Nevertheless, I~recognize that code\-breakers at Bletchley Park during
the \mbox{Second} World War were definitely on the side of good.
We~all know about the prowess of Alan \mbox{Turing}, who played a key role at the
routine (this word is too strong) decryption of the German Enigma cipher~\cite{Turing}.
But who remembers \mbox{Marian} \mbox{Rejewski}, who actually used pure
(and beautiful) mathematics to break Enigma with two \mbox{colleagues}
\emph{before} the War even started?~\cite{Rejewski}
Indeed, who remembers \mbox{except} yours truly and nation\-al\-istic Poles such
as my friend Artur Ekert?
\mbox{Certainly} not filmmakers!~\cite{imitation}
And who remembers William Tutte, who broke the much more difficult Lorenz cipher
(code\-named Tunny by the Allies),
which allowed us to probe the mind of Hitler?~\cite{Tutte}
Tutte moved on to found the Computer Science department at the University of Waterloo,
Canada, now home of IQC, the Institute for Quantum Computation,
but never said a word until the 1990s about how he won the War for us~\cite{Fish}.
The Canadian Communications Security Establishment 
pays homage with its Tutte Institute for Mathematics and Computing.
But~who else remembers those silent heroes on the code\-breaking side?
I~am getting \mbox{carried} away by emotions as I~type these words while
flying from T\={o}ky\={o} to Calgary, on my way home after the amazingly successful
5th Annual Conference on Quantum Cryptography, QCrypt 2015~\cite{QCrypt}.

Regardless of the side to which good belongs, the obvious question is:
\emph{Who will win the battle} between code\-makers and code\-breakers?
More specifically, how do the recent advances in Quantum Information Science (QIS)
change this age-old issue? Until the mid-twentieth century, History has taught us
that code\-makers, no matter how smart,
have been systematically outsmarted by code\-breakers, but it ain't always been easy.
For~instance, \emph{le chiffre ind\'e\-chif\-frable}, \mbox{usually} \mbox{attributed} to Blaise de Vigen\`ere
in~1585, but actually invented by Giovan Batista Belaso 32 years earlier,
remained invulnerable until broken by Charles Babbage 
more than three centuries after its invention.
(Baggage is best known for the \emph{Analytical Engine},
which would have been the first programmable computer had the technology
of his days been able to rise up to the challenge.)
The~apparent upper hand of code\-breakers, despite the still enduring invulnerability
of the \emph{chiffre ind\'e\-chif\-frable}, prompted American novelist and high-level
amateur cryptanalyst Edgar Allan Poe to confi\-dently \mbox{declare} in 1841 that
``It~may be roundly asserted that \mbox{human} ingenuity cannot concoct a cipher which human ingenuity cannot \mbox{resolve}''~\cite{Poe}.
Poe,~do I~need to mention, was among other things the author of
\emph{The Gold-Bug}~\cite{GoldBug}, published in June 1843.
This extraordinary short story centring on the decryp\-tion of a secret \mbox{message}
was instrumental on kindling the career of prominent cryptographers, such as
William Friedman's, America's foremost cryptanalyst of a bygone era,
who read it as a child~\cite[p.~146]{Rosenheim}.

Cryptography was set on a firm scientific basis by Claude Shannon,
the father of information theory~\cite{ShannonIT}, as the first half of the twentieth century
was coming to a close~\cite{ShannonCrypt}. Actually, it's likely that his groundbreaking work was
achieved several years earlier but kept classified due to the War effort.
In~any case, Shannon's theory was resolutely set in the context of
\emph{classical physics}. In~retrospect, this is odd since it was clearly
established at that time that Nature is ruled not by the Laws envisioned
centuries earlier by Sir Isaac Newton, and not even by those more modern
of Albert Einstein, but by the counterintuitive features of the emerging
quantum mechanics. Shannon was well aware of this revolution in physics,
but he probably did not think it relevant to the {foundations} of
information theory, which he developed as a purely abstract theory.

In~particular, Shannon did not question the ``fact'' that 
encrypted infor\-mation transmitted from a sender (code\-named \mbox{Alice})
to a receiver (code\-named Bob) could be copied by an eavesdropper
(code\-named Eve) without causing any disturbance noticeable by \mbox{Alice} and Bob.
From this unfounded assumption, Shannon proved a famous theorem according
to which perfect secrecy requires the availability of a shared secret key as long as
the message that \mbox{Alice} wishes to transmit securely to Bob, or more precisely
as long as the \emph{entropy} of that message,
and that this key cannot be reused~\cite{ShannonCrypt}.
This theorem is mathematically impeccable, but it is nevertheless irrelevant
in our quantum-mechanical world since the assumption
on which its proof is based does not hold.

My purpose is to investigate the issue of whether or not Poe was right in his
sweeping mid-nineteenth century statement.
Could it be indeed that code\-breakers will continue to have the upper hand over
code\-makers for the rest of eternity?

\section{The Case of Classical Code\-makers against Classical Code\-breakers}

The first electronic computers were designed and built to implement
Tutte's beautiful mathematical theory on how to break the high-level
German code during World War~II.
They were code\-named the \emph{Colossus} and ten of them were built
in Bletchley Park~\cite{Colossus}.
As mentioned in the Introduction, they were instrumental in allowing us to win the~War.
However, in order to secure secrecy of the entire Bletchley Park operation,
they were smashed to bits
(funny expression when it concerns computers!)\ once the War was over.
Consequently, I~``learned'' as a child that the first electronic
computer in history had been the American \textsc{Eniac}, when in fact it was the eleventh!
Little did the pioneers of the Colossus imagine that, by an ironic
twist of fate, they had unleashed the computing power that was to bring
(temporary?)\ victory to the code\-makers.
In~a sense, code\-breakers had been the midwife of the instrument
of their own destruction. Perhaps.
\mbox{Indeed}, the rise of \emph{public-key cryptography} in the 1970s
had led us to believe that an increase in computing power could only be in favour
of code\-makers, hence at the detriment of code\-breakers.

But well before all this took place,
a~cryptographic method that offers perfect secrecy, which later came to be known as the
\emph{one-time pad}, had already been invented
in the nineteenth century. It~is usually attributed to Gilbert Vernam, who was
granted a US Patent in 1919~\cite{Vernam}.
However, according to prime historian David Kahn, Vernam had not realized
the crucial importance of never using the same key twice until Joseph Mauborgne
pointed it~out~\cite[p.~398]{Kahn}. But~it was later discovered that the
one-time pad had been invented 35 years earlier by Frank Miller, a Sacramento
banker~\cite{Miller}. Its~perfect security was demonstrated subsequently by
Shannon~\cite{ShannonCrypt}.
In~any case, the one-time pad requires a secret key as long as
the message to be transmitted,
which makes it of limited practical use.
It~was nevertheless used in real life, for instance on the red telephone
between John Kennedy and Nikita Khrushchev during the Cold~War~\cite{RedTelephone},
as well as between Fidel Castro and Che Guevara after the latter had
left Cuba for Bolivia~\cite{Che}.
But~in our current information-driven society, we need a process by which
any two citizens can enjoy confidential communication.
For~this, a method to establish a shared secret key is required.
Could this be achieved through an \emph{authentic public channel},
which offers no protection against eavesdropping?

The first breakthrough in the academic world came to Ralph Merkle in 1974,
who designed a scheme capable of providing a quadratic advantage to
code\-makers over code\-breakers.
Merkle's scheme is secure under the sole
assumption (still unproven to this day) that some problems can only
be solved by exhaustive search over their space of potential solutions.
At~the time, Merkle was a graduate student at the University of California
in Berkeley, enrolled in a computer security class.
Unable to make his ideas understood by his professor,
Merkle ``dropped the course, but kept working on the idea''~\cite{CS244}.
After several years, he prevailed and his landmark paper was
finally published~\cite{Merkle}.
However, Whitfield Diffie, a graduate student ``next door'',
at Stanford University, had similar ideas independently, albeit shortly after Merkle.
But~Diffie was lucky enough to have an advisor, Martin Hellman,
who understood the genius of his student.
\mbox{Together}, they made the concepts of public-key cryptography and digital signature
\mbox{immensely} popular~\cite{DH}, two years before Merkle's publication.

A~few years later, Ronald Rivest, Adi Shamir and Leonard Adleman,
inspired by the Diffie-Hellman breakthrough, proposed an implementation
of public-key cryptography and digital signatures
that became known to all as the RSA crypto\-system~\cite{RSA}.
And thus, history was made.
The~fact that the RSA crypto\-system had in fact been invented in 1973
by Clifford Cocks~\cite{Cocks},
at the British secret services known as GCHQ, is of little relevance to the
practical importance of the discovery on what was to become the Internet.
As~long as the factorization of large numbers remained infeasible,
the code\-makers had finally won the battle, proving Poe wrong.
Soon, electronic safety all over the Internet revolved around this RSA crypto\-system,
as well as the earlier invention known as the Diffie-Hellman key establishment protocol~\cite{DH}.
At~about the same time, Robert McEliece invented another approach,
based on error-correction codes~\cite{McEliece}, which did not come into practical use
because it required much longer keys than \mbox{either} the RSA or the Diffie-Hellman solution.
Later, the same apparent level of security was obtained with significantly shorter keys
by bringing in the number-theoretic notion of elliptic curves~\cite{ECC1,ECC2}.
And~the Internet was a happy place.
Or~so it seemed.
\looseness=-1

End of story?

\section{The Unfair but Realistic Case of Classical Code\-makers against Quantum Code\-breakers}

End of story?
Not quite!
In~the early 1980s, Richard Feynman~\cite{Feynman82,Feynman85} and,
independently, David Deutsch~\cite{Deutsch},
invented the theoretic notion of a \emph{quantum computer}.
This hypothetical device would use the counterintuitive features of
quantum mechanics for computational purposes.
At~first, it was not clear that quantum computers, even if they could be built,
could speed up calculations.

And~then, in 1994, Peter Shor~\cite{Shor}, and independently
Alexis \mbox{Kitaev}~\cite{Kitaev}, discovered that
quantum computers have the power to factor large numbers and extract
discrete logarithms efficiently, bringing to their knees not only the RSA crypto\-system
but also the Diffie-Hellman key establishment scheme, even if based on elliptic curves.
As~a society, we are extremely fortunate that Shor's and \mbox{Kitaev}'s discoveries were
made before a quantum computer had already been built for some other
purposes (such as computational physics and chemistry).
Quite literally, this saved civilization from catastrophic collapse.
But~now that we have known about the looming threat for over two decades,
surely we are active at deploying solutions that have at least a fighting chance
to withstand the onslaught of a quantum computer.

Well, not really. \verb+:-(+

The general apathy towards the quantum threat to worldwide security on the Internet
and beyond is quite simply appalling. Why react today (or~more appropriately
twenty years ago) when we can quietly wait for disaster?
After all, no serious business model looks more than five years in the future,
and it \emph{would} be expensive to change the current cryptographic infrastructure.
And~\mbox{indeed}, a full-scale quantum computer is unlikely to materialize in the next five years.
Except perhaps in an ultra-secret basement somewhere, be it governmental
of industrial\ldots{}
But when (not~``if{\small \,}'') this happens, all \emph{past} communications will become
insecure to whomever was wise
enough to have stored the Internet traffic that was until then undecipherable.
The~fact that current cryptographic techniques are susceptible to
being broken \emph{retroactively}
is their main conceptual weakness.
Any~secret entrusted to them today, even if it is indeed currently \mbox{secure} (something that
we do not know how to prove), will be exposed as soon as a sufficiently large
quantum computer becomes operational.

So, was Poe right after all?
Are code\-breakers poised to regain their \mbox{upper} hand?
Not necessarily!
Alternative encryption methods have been designed, which are not (yet) known to be vulnerable
to a quantum attack, ironically \mbox{including} the historical McEliece approach~\cite{McEliece},
which had been scorned upon its invention because of the length of its~keys.
More recent approaches based on hash functions, short vectors in lattices and
multivariate polynomials are \mbox{being} vigorously investigated.
The emerging field of \emph{post-quantum cryptography}
is \mbox{devoted} to the study of (hopefully) quantum-resistant encryption~\cite{Post-Quantum1,Post-Quantum2}.
Unfor\-tu\-nately, we~cannot prove that any of these alternatives is secure,
but at least they are not already known to be compromised by the advent
of a quantum computer.
Well, in the case of lattice-based cryptography~\cite{lattice},
this is not so clear anymore~\cite{tricky,cautionary,Biasse}.
But~one thing is sure: we cannot hope to be protected by these techniques
if we don't use them!
On~the other hand, some of these more recent schemes could in fact be
\emph{less secure} than RSA against a \emph{classical} attack, simply because
they have not yet stood the test of time.
Therefore, a transition to these new techniques should be carried out
with the utmost care. But it \emph{must} be carried~out.
\looseness=-1

Michele Mosca likes to tell the following tale.
Let~$x$ denote the length of time (in~years) that you want your secrets to remain secret.
Let~$y$ denote the time it will take to \mbox{re-tool} the current
infrastructure with quantum-safe encryption (\mbox{assuming} that such a thing actually exists).
Let~$z$ denote the time it will take before a full-scale quantum computer is operational.
Mosca's ``theorem'' tells us that if \mbox{$x+y>z$}, then it is time to panic!
Sadly, it may even be that \mbox{$y>z$}, meaning that it's already too late to
avoid a complete meltdown of the Internet.
So, \emph{what are we waiting~for?}

It turns out that the American National Security Agency (NSA) is taking this threat
\emph{very} seriously indeed. This last August (2015), they issued a directive called
``Cryptography Today'' in which they announced that they
``will initiate a transition to quantum resistant algorithms in the not too distant future''~\cite{NSA}.
Most significantly, they wrote: 
``For~those partners and vendors that have not yet made the transition to Suite~B elliptic curve algorithms, we recommend not making a significant expenditure to do so at this point but instead
to prepare for the upcoming quantum resistant algorithm transition''.
Said plainly, even though elliptic-curve cryptography is believed to be more secure than
first-generation public key solutions against classical cryptanalysis,
it is no longer considered to offer sufficient long-term security under the looming threat
of a quantum computer to be worth implementing at this point.
It's nice to see that \emph{someone} is paying attention.
For~once, I'm glad that the NSA is listening!  \verb+:-)+

From a theoretical perspective, despite what I~wrote above,
it~\emph{is} possible to have provably quantum-safe encryption under the so-called
random oracle model, which is essentially the model that was used by Merkle in his
original 1974 invention of public key establishment~\cite{CS244}.
In~a classical world, this model roughly corresponds to the assumption that there are
problems that can only be solved by exhaustive search over their space of potential solutions.
In~the quantum setting, exhaustive search can be replaced by a celebrated algorithm
due to Lov Grover, which offers a quadratic speedup~\cite{Grover},
but no more~\cite{BBBV}.

Recall that Merkle's original idea brought a quadratic advantage to code\-makers over code\-breakers.
But since Grover's algorithm offers a quadratic speedup to code\-breakers,
this completely offsets the code\-makers' advantage.
As~a result, code\-breakers can find the key established by code\-makers
in the same time it took to establish~it!~\cite{ICQNM}
The obvious reaction is to let the code\-makers use quantum powers as well,
but please remember that in this section, we consider quantum code\-breakers but only
classical code\-makers. Nevertheless, I~have discovered with 
\mbox{Peter} \mbox{H{\o}yer}, \mbox{Kassem} \mbox{Kalach},
\mbox{Marc} \mbox{Kaplan},
\mbox{Sophie} \mbox{Laplante} and \mbox{Louis} \mbox{Salvail}
that Merkle's idea can be modified in a way that if the code\-makers are willing to
expend an effort proportional to some parameter $N\!$, they can obtain a shared key
that cannot be discovered by a quantum code\-breaker who is not willing to expend an effort
proportional to~$N^{7/6}$~\cite{Crypto2011}.
As~I~said, this is purely theoretical because it is not possible to argue that such an
advantage offers practical security.
Indeed, $N$ would have to be astronomical before a key that is obtained in, say,
one second would require more than one year of code\-breaking work.
In~contrast, Merkle's quadratic advantage is significant for reasonably small
values of~$N\!$.
Nevertheless, our work should be seen as a proof of principle.
Now that we know that \emph{some} security is possible in the
unfair case of classical code\-makers against quantum code\-breakers,
it is worth trying to do better
(or~prove that it is not possible).

Coming back to the question asked at the end of the Abstract,
quantum mechanics appears to be a curse for the protection of privacy in this unfair context,
which is hardly surprising since only code\-breakers were assumed to use~it!

\section{Allowing Code\-makers to Use Quantum Computation}

The previous section considered a realistic scenario in which simple citizens want to
protect their information against a much more powerful adversary.
Indeed, it is likely that quantum computers will initially be available
only to large governmental, industrial and criminal organizations.
Furthermore, it is safe cryptographic practice to assume that your adversary
is computationally more powerful (and possibly also more clever) than you~are.

Nevertheless, in the more distant future, one can imagine a world in which
quantum computers are as ubiquitous as classical computers are today.
When this happens, code\-makers will no longer be limited to classical computing.
Can~this restore the balance?
Or~even better, could the availability of quantum computers turn out to be to the
advantage of code\-makers, just as had been the availability of ever increasing classical
computational power since the inception of public-key cryptography in the mid-1970s?
Unfortunately, I~am not aware of any encryption technique that would benefit from
quantum computation sufficiently to offset the benefits that quantum computation
would bestow on code\-breakers.  

For instance, it~is easy to \emph{partially} repair Merkle's approach~\cite{ICQNM}
if the code\-makers are also allowed to use
Grover's algorithm, or more precisely a variant known as BBHT~\cite{BBHT}.
Having expended an effort proportional to~$N$ in order to obtain a shared key,
they can create a puzzle on which classical code\-breakers would have to
expend an effort proportional to $N^3$, a~clear improvement over the quadratic
advantage of the original classical Merkle approach.
However, a quantum code\-breaker would simply use Grover's algorithm
to obtain the key after an effort proportional to~$N^{3/2}$.
This is not a complete break,
but this quantum scheme is not as secure as Merkle's original would have been
against a classical adversary.
So,~we see that quantum-mechanical powers have helped the code\-breakers
more than the code\-makers.
Can~code\-makers use quantum powers in a more clever manner?
Well, we have developed a less obvious Merkle-like quantum key establishment scheme
against which a quantum code\-breaker needs to spend a time proportional to~$N^{7/4}$~\cite{Crypto2011}.
This is still not quite the quadratic advantage that was possible in an all-classical world,
but it is reasonably close and possibly secure enough to be used in practice.

Nevertheless, quantum mechanics still appears to be a curse for the protection of privacy
even when code\-makers are also allowed to make use of~it.

\section{Allowing Code\-makers to Use Quantum Communication}

Until now, we had restricted all communication between code\-makers to be classical.
It~turns out that quantum communication comes with a great advantage because of the
no-cloning theorem~\cite{NoCloning},
which says that the state of elementary particles cannot be copied even in principle.
This is \emph{precisely} what causes the demise of the ``famous'' theorem by Shannon
mentioned at the end of the Intro\-duc\-tion.
Quantum information transmitted between code\-makers
can\emph{not} be copied by an eavesdropper without causing a detectable disturbance.

Inspired by an unpublished manuscript written by Steven Wiesner in April 1968, while
he was participating in the Columbia University student protests~\cite{Wiesner},
Charles Bennett and~I realized in 1982 that quantum mechanics provides us with
a channel on which passive eavesdropping is impossible.
This led us and Seth Breidbart to a write down what would become
the leitmotif of the nascent field of \emph{quantum cryptography}.
\begin{quote}
\textsf{When elementary quantum systems, such as
polarized photons, are used to transmit digital information,
the uncertainty principle gives rise to novel cryptographic
phenomena unachievable with traditional transmission
media, e.g.~a communications channel on which it is
impossible in principle to eavesdrop without a high probability
of being detected.}~\cite{BBB82}
\end{quote}
Armed with this idea, we devised a cryptographic protocol in which
a one-time pad could be safely reused indefinitely,
as long as no eavesdropping is detected.
This secure reuse of a one-time pad
is precisely what Shannon had mathematically demonstrated to be impossible:
all security is lost as soon a ``one-time'' pad is used twice.
Our advantage, of course, comes from the fact that \emph{we} could detect eavesdropping
and discontinue the use of a pad as soon as it had been compromised
(yet~providing perfect secrecy even on the last message that was sent),
whereas \emph{he} had no fundamental way to detect eavesdropping,
and therefore he was forced to play safe.

In~more detail, Shannon proved that the one-time pad is unconditionally \mbox{secure}
provided the shared key is perfectly random, completely unknown of the
eavesdropper, and used once only.
However, even though no information leaks concerning the
message in case of interception, information \emph{would} leak concerning the
key itself. This is of no consequence as long as the key is never reused.
But~if it is, the key-secrecy condition is no longer fulfilled the second time,
which is why the system becomes insecure.
It~follows that a ``one-time'' pad \emph{can} be reused safely,
Shannon's theorem notwithstanding, provided the
previous communications have not been subject to eavesdropping, and it remains
secure the first time that it~is.

Expounding on these ideas, we wrote our paper on
``How to re-use a one-time pad safely'' in 1982
and had it published\ldots{} a few months ago, $2^5$~years later!~\cite{BBB82}.
The~reason it took so long to publish is that as soon as it was about to be
rejected from the \emph{Fifteenth Annual ACM Symposium on Theory of Computing},
Bennett and~I had a much better idea:
we~realized that it is more practical to use the quantum channel
to establish a shared secret random key,
and then use this key as a \emph{classical}
one-time pad to encode the actual message,
rather than use the channel to transmit the message directly.
The main advantage of this indirect approach
is that even if most of the quantum information is
lost in the channel---indeed, optical fibres are not very transparent
to single photons over several kilometres---a random subset of a random key
is still a (shorter) random key.
In~contrast, a small random subset of a meaningful message is fairly likely
to be mostly random and totally useless.

Thus was born \emph{Quantum Key Distribution}, which is now called simply~QKD.
We~presented QKD for the first time at the 1983 IEEE International Symposium
on Information Theory~\cite{St-Jovite}, but each paper was allowed only a one-page
abstract. Consequently, our protocol had to wait
another year before it could be published in the Proceedings of a conference
held in Bengal\={u}ru, India, where I~had been invited to present any paper
of my choice~\cite{BB84}.
I~suspected that the idea of QKD was likely to be rejected if submitted to a conference with
full published proceedings, which is why
I~seized the opportunity provided by a blank-cheque invitation to sneak it
at that conference!
This is how our original QKD protocol came to be known as ``BB84'', where the
Bs stand for the authors, despite the fact that we had invented and presented it
in~1983.
Thirty years later, \emph{Natural Computing} (Springer)
and \emph{Theoretical Computer Science} (Elsevier)
\mbox{decided} to join forces and publish special BB84 commemorative issues.
This is how the earlier 1982 paper came to be published~\cite{BBB82},
whereas the original ``BB84~paper'' was published for the first time in a journal~\cite{BB84TCS}.
For~more information on the early history of quantum cryptography, please
read Ref.~\cite{Awaji}.

It was fairly easy to show that BB84 is secure against the most obvious
\mbox{attacks} that an eavesdropper might attempt~\cite{BBBSS}.
However, it took ten years after its invention before a complete formal proof of
unconditional security, taking into account \emph{any} attack possible according
to the laws of quantum mechanics, was obtained~\cite{Mayers}.
Well, not exactly.
This early proof, as well as the few that followed for the purpose of simplifying~it,
contained a major oversight. They proved that the key established by BB84
(and~other similar QKD protocols) was perfectly secret\ldots{}
provided it is never used!
Indeed, \mbox{Renato} \mbox{Renner} and Robert K\"onig realized ten years later that a clever adversary
could keep the eavesdropped information at the quantum level
(unmeasured). Later, when the key is used,
say as one-time pad, the information that it leaks on the key
(which would not be a problem in classical cryptography
since the key would not be reused) could inform the eavesdropper about the
appropriate measurement to make in order to learn more of the key and,
therefore something about the message itself~\cite{Renato}.
At~first, this was only a theoretical worry, but then it was shown that
the danger is real because one could purposely design a QKD scheme
that could be proved secure under the old definition, but that really leaked
information if the ``secret'' key is used~\cite{Koenig}.
Fortunately, the adequate (``composable'') definition was given and BB84
was correctly proven secure a few months later~\cite{BB84secure}. 

Et voil\`a!
Quantum cryptography offers an unbreakable method for code\-makers
to win the battle once and for all against any possible attack available
to code\-breakers, short of violating the widely accepted laws of physics.
Despite the discouraging news brought about by the previous sections,
in which quantum mechanics appeared to be a curse for code\-makers,
in the end it is a blessing for the protection of privacy.

As my much missed dear friend Asher Peres once said,
``\textsl{The quantum taketh away and the quantum giveth back}''.
Indeed, quantum mechanics can be \mbox{exploited} to break the cryptography that is currently
deployed over the worldwide Internet, via Shor's algorithm, but 
quantum mechanics has also provided us with the ultimately secure solution.
(To~be historically exact, the quantum giveth ``back'' ten years \emph{before}
it taketh away!)

Poe~was wrong. End of story!

Oh well\ldots{} Not so fast.
Poe was wrong \emph{in theory}.
Now, one has to build an apparatus that implements QKD as specified by
the theoretical protocol.
Exactly?
Not possible!
Any real implementation will be at best an approximation of the ideal protocol.
The first prototype was built by Bennett and me, with the help of three
students (two of whom have become highly respected researchers in the field)
as early as~1989, even though the journal paper was published
a few years later~\cite{BBBSS,SciAm}.
This prototype was not intended to be more than a proof of principle and
some of its parts made such loud noises that we could literally hear the bits
fly~by\ldots{} and zeroes did not make the same noise as ones.
So,~this first implementation was secure provided the eavesdropper is deaf!

Afterwards, serious experimental physicists entered the game and
ever increas\-ingly sophisticated devices have been built, capable of establishing
\mbox{secret} keys over longer and longer distances. This business became so serious
that companies sprung up to market QKD equipment, such as
ID~Quantique~\cite{idQ} in Switzerland. 
China has recently announced that it has almost completed the installation
of a quantum communications network stretching two thousand kilometres
from Beijing to Shanghai~\cite{China}.
Several countries have plans to move the quantum highway to space,
so that distances will no longer be an issue.

In~the mean time, a new breed of (typically friendly) \mbox{pirates} has sprung~up:
the \emph{Quantum Hackers}.
In~2009, a team lead by Vadim Makarov completed a
``full-field imple\-men\-ta\-tion of a complete attack on a running QKD connection; an~installed eavesdropper obtained the entire `secret' key, while none of the param\-eters monitored by the legitimate parties indicated a security breach''~\cite{Makarov}.
Of~course, this was not an attack against BB84 or any other provably secure
QKD protocol, which would have been an attack against quantum mechanics itself:
this was an attack against one particular imperfect \emph{implementation} of a perfect idea.
The specific flaw was eradicated\ldots{} and Makarov found another weakness!

And so, the game of cat and mouse between code\-makers and code\-breakers
continues. Only the battlefield has shifted from the realm of mathematics and
computer science to the realm of physics and engineering.
Nevertheless, even an imperfect implementation of QKD has a significant
advantage over classical systems: it~\emph{must} be attacked while the
key establishment process is taking place. There is nothing to store for
subsequent code\-breaking when new technology or new algorithms
become available. If~the technology is available today for the implementation
of some imperfect version of QKD but not yet for breaking~it,
everlasting security is achievable.
Similarly, I~have not mentioned the fact that the deployment of QKD requires
the availability of an authenticated classical channel between the code\-makers
to avoid a person-in-the-middle attack, much as was the case
for Merkle's classical approach in 1974. However, if the code\-makers can establish
short-lived secure authentication keys by any method,
those keys can give rise to everlasting security through the use of~QKD,
again an advantage that has no classical counterpart~\cite{Unruh}.

Nevertheless, it is legitimate to wonder if there is any hope of one day building an
implementation of QKD so close
to the ideal protocol that it will effectively be secure against all possible attacks,
regardless of the code\-breaker's technology and computing time?
It~is tempting to say that this would be Mission: Impossible.
Surely, an army of Makarovs will spring up with increasingly clever ideas
to defeat increasingly sophisticated (yet~imperfect) implementations of~QKD.
Said otherwise, surely Poe was right in the~end.

Well\ldots{} Maybe not!
A~new approach to QKD has sprung~up, based on a brilliant idea put forward
by Artur Ekert as early as 1991~\cite{E91}.
Instead of basing the security of QKD on the impossibility of cloning
quantum information---more fundamentally the impossibility of obtaining
classical information on a quantum system without
disturbing~it~\cite{BBM92}---Ekert's idea was to base the security
of QKD on violations of Bell inequalities~\cite{Bell}
in entangled nonlocal quantum systems~\cite{EPR}.
Even though Ekert's original 1991 QKD protocol cannot give rise to an apparatus that
would be more secure than one based on BB84~\cite{BBM92}, his fundamentally revolutionary idea
opened the door to other theoretical QKD protocols that have the potential to be
secure \emph{even if implemented imperfectly}.
The~security of those so-called ``device-independent QKD protocols''
would depend only on the belief that information cannot travel faster than light,
that the code\-makers are capable of choosing their own independent randomness,
and of course that they live in secure private spaces
(since there is no need for code\-breakers if the adversary is capable
to physically eavesdrop over the code\-makers' shoulders!).
In~the extreme case, highly theoretical device-independent QKD protocols
have been designed whose security does not even depend on the validity
of quantum mechanics itself!
A~recent survey of this approach is found in Ref.~\cite{ultimate}.

The catch is that the implementation of fully device-independent QKD protocols
represents formidable technological challenges. It~is not clear that we shall
ever reach the required sophistication to turn this dream into reality.
Nevertheless, a first essential step towards this goal has been achieved very recently
by Ronald Hanson and collaborators in the Netherlands when they performed
a long-awaited experiment in which they closed both the locality and the
detection loopholes in experimental violations of Bell inequalities~\cite{Hanson,LoopholeFree}.

Shall we ever be able to build such a device?
If~so, the code\-makers will have the final laugh.
But what if not?

Was Poe right in the end? The jury is still out!

\subsubsection*{Acknowledgments.} 

I~am grateful to all those with whom I~have had fruitful discussions on these issues
in the past 36 years, starting with my lifelong collaborators Charles Bennett
and Claude Cr\'epeau. I~thank Michele Mosca for allowing me to quote his ``theorem''.
I~am also grateful to R\={u}si\c{n}\v{s} Freivalds for his invitation
to present this paper to this 42nd International Conference on Current Trends
in Theory and Practice of Computer Science (SOFSEM) and for his
involvement in my 1998 election as Foreign Member of the Latvian
Academy of Sciences.
This work was supported in part by Canada's Natural
Sciences and Engi\-neering Research Council of Canada (\textsc{Nserc}),
the Institut transdisciplinaire d'informatique quantique (\textsc{Intriq}),
the Canada Research Chair program
and
the Canadian Institute for Advanced Research (\textsc{Cifar}).

\end{document}